# Cooperating epidemics of foodborne diseases with diverse trade networks

Yong Min[1], Ying Ge[1], Xiaogang Jin[2], Jie Chang[1]

[1] College of Life Sciences, Zhejiang University, Hangzhou 310058, China

[2] College of Computer Science, Zhejiang University, Hangzhou 310027, China

**Abstract**

The frequent outbreak of severe foodborne diseases warns of a potential threat that the global trade networks could spread fatal pathogens. The global trade network is a typical overlay network, which compounds multiple standalone trade networks representing the transmission of a single product and connecting the same set of countries and territories through their own set of trade interactions. Although the epidemic dynamic implications of overlay networks have been debated in recent studies, some general answers for the overlay of multiple and diverse standalone networks remain elusive, especially the relationship between the heterogeneity and diversity of a set of standalone networks and the behavior of the overlay network. In this paper, we establish a general analysis framework for multiple overlay networks based on diversity theory. The framework could reveal the critical epidemic mechanisms beyond overlay processes. Applying the framework to global trade networks, we found that, although the distribution of connectivity of standalone trade networks was highly heterogeneous, epidemic behavior on overlay networks is more dependent on cooperation among standalone trade networks rather than on a few high-connectivity networks as the general property of complex systems with heterogeneous distribution. Moreover, the analysis of overlay trade networks related to 7 real pathogens also suggested that epidemic behavior is not controlled by high-connectivity goods but that the actual compound mode of overlay trade networks plays a critical role in spreading pathogens. Finally, we study the influence of cooperation mechanisms on the stability of overlay networks and on the control of global epidemics. The framework provides a general tool to study different problems on overlay networks.

**Introduction**

For decades, scientists from various fields have sought to understand epidemic dynamics on complex networks (Pastor-Satorras and Vespignani, 2001; Newman, 2002; Hufnagel et al., 2004; Colizza et al., 2006; 2007; Meloni et al., 2009). Recently, the frequent outbreak of trade-mediated foodborne diseases has posed a challenge to the general approach of epidemic dynamics on the network (Todd, 2008; Osterholm, 2011; Kupferschmidt, 2011a; 2011b). The spread of trade-mediated pathogens always involves multiple vectors, which interact with each other to form an overlay network (Funk and Jansen, 2010; Marceau et al., 2011; Wang et al., 2011; Dickison et al., 2012). Nevertheless, the classical approach of epidemic dynamics on the networks does not consider the interaction among multiple networks and networks with multiple types of edges.

An overlay network is the integration of two or more standalone networks connecting the same set of nodes with their own set of edges (Fig. 1) (Funk and Jansen, 2010; Marceau et al., 2011). The interest in overlay networks is not only to restrict the spread of pathogens in the biological sense, but also to spread awareness or influence social networks and the movement of individuals in transportation networks (Funk and Jansen, 2010; Marceau et al., 2011). For example, mobile phone viruses always involve different spreading routes related to different media. Moreover, with the rapid development of information technology, all social contact among individuals is usually dependent on various interacting tools, including direct talk, email, instant messaging, and social networking services. Therefore, the natural next step was to understand the consequences of overlay processes.

The earlier studies, however, used simple models for two overlay networks, calling into question their applicability to the more complex multiple overlay networks that existed in the real world (Funk and Jansen, 2010). A similar problem also exists in the study of multiple networks and interdependent networks, which also model the interaction among different networks (Buldyrev et al., 2010; Parshani et al., 2010a; 2010b; Gao et al., 2012; Paczuski et al., 2012). A more recent study attempted to address this concern and provided a mathematical method to deal with epidemic dynamics on any number of interdependent networks (Paczuski et al., 2012). Nevertheless, two critical questions remain: (i) How general are the methods to real overlay networks, and (ii) how to identify and quantify the mechanisms beyond the overlay processes.

Here, we rigorously examine the relationship between overlay diversity and its epidemic dynamics. Specially, we answer the question of how and why the percolating cluster of the overlay networks is related to whether the set of standalone networks is larger or smaller (richness) and more or less different (evenness). First, we establish a general diversity analysis framework to describe and analyze overlay networks. Second, we reveal the key mechanism affecting epidemic dynamics on worldwide trade networks. Third, based on random and targeted perturbation scenarios, we also discuss the stability of the overlay networks for providing references to design rational and effective immunization strategies. Finally, we analyze the overlay networks related to the spread of 7 real pathogens to support the conclusions from simulation studies.

**Materials and methods**

Food and Agriculture Organization of the United Nations (FAO, http://faostat.fao.org) provides the world list of 254 countries and territories with 574 traded agricultural products. Because vegetables and fruits have been reported as the most probable vector of recent outbreaks of foodborne diseases, we focus on the vegetable and fruit products and their trades. A standalone trade network describes the trade of single vegetable or fruit product, and is an unweighted directed network, whose nodes represent countries and territories and whose edges represent primary trade interaction between two nodes. In order to get standalone trade networks from raw FAO data, we compile the original data by three steps.

1) Combining items. FAO classify the trade items according to its product, form and processing methods. For example, a single product, apple, is divided into fresh apple, concentrated apple juice and single strength apple juice. In order to reflect the trade of single product, we combine all items of one product into a standalone trade network. Another reason to perform combination operation is that cross infection always occur with two products but nothing on the form and processing methods. The detail of combination is listed in the *supplementary information*. If we combine items into a standalone trade network, we set $Q_{ij} = \sum_s Q_{ij}^s$, where $Q_{ij}$ is the trade quantity between node *i* and *j* in the standalone network, and $Q_{ij}^s$ is the trade quantity of item *s* between node *i* and *j*. The unit of trade quantity is ton according to the FAO standard.

2) Filtering primary edges. Because the trade networks are high density in connection, we delete minor edges ($Q_{ij} \leq 1,000$ tons) and maintain primary edges ($Q_{ij} > 1000$ tons). The resulted set of standalone trade networks account for 95% global trade quantity of vegetables and fruits, but only involve 15% of all trade interactions. Therefore, the result standalone networks are the stem of global vegetable and fruit trades, and could reflect the major features of global trade networks.

3) Unweighting networks. Because we have classified the edges of all standalone networks into primary and minor edges, we treat all primary edge as the same capacity in spreading pathogens, i.e. do not consider their trade quantity. The simplification is convenient for demonstrate our diversity analysis framework at the first time.

By compiling original data from FAO, we got 83 standalone trade networks. Each network represents the trade of single type good (e.g. potatoes, cassava or apples). The resulting standalone trade networks are graph comprising 202 nodes, and 1 to 1893 unweighted direct edges. These standalone networks are used to integrate into overlay networks for studying diversity effect.

**Results and discussion**

*Diversity analysis framework of epidemics on overlay networks*

The framework consists of three components: 1) the definition of overlay diversity, including overlay richness and evenness, 2) the declaration of functioning and stability of overlay networks, and 3) the partition of overall diversity effects into different mechanisms.

A system of multiple overlay networks is based on a set of standalone networks N={$\Gamma_1=(V,E_1),\Gamma_2=(V,E_2),…,\Gamma_n=(V,E_n)$}, which connect the same set of nodes, $V$, with different sets of edges, $E_k$. ($1 \leq k \leq n$) (Fig. 1). Nodes represent agents for infection (e.g., countries and territories in trade networks or individuals in social networks), while edges correspond to potential transmission routes between pairs of infecting agents. The overlay network, $\Gamma_O=(V,E_O)$, is the combination of these standalone networks, where $E_O = E_1 \cup E_2 \cup … \cup E_n$, that is, if there is edge $e_{ij}$ in at least one $\Gamma_k$, the edge exists in the overlay network (Fig. 1). According to the rules, any standalone network could be treated as the overlay network of itself.

In order to understand the effects that changes in network diversity will have on epidemic dynamics, it is important to define some metrics for diversity. Diversity is not easily defined, but may be thought of as the richness and evenness of overlay networks. Overlay richness ($R$) is defined as the number of standalone networks used to form the overlay network, whereas overlay evenness ($D$) quantifies similarity among standalone networks, and is defined as:

$$D = \max_{k,k' \in N}(H_{kk'})$$

where $H_{kk'}$ is the Hamming distance between two adjacency matrixes of standalone network $k$ and $k'$, which is the number of positions at which the corresponding symbols (0 for no trade and 1 for two nodes linked by a trade) are different. Based on the richness and evenness, we could quantify how overlay diversity could affect the epidemic dynamics and stability of overlay networks.

In order to describe epidemic dynamics, we use edge percolation processes to model epidemic spreading in the network (Newman, 2003; Parshani et al., 2010a). In the process, each edge becomes active with a given percolating probability, $p$, and all active edges could lead to one or more weakly connected percolating clusters in directed networks (a weakly connected component is a maximal group of nodes that are mutually reachable by violating the edge directions). The number of nodes or edges involved in the component could measure the size of these percolating clusters. Here, we focus on the maximal percolating cluster (MPC) and use it to indicate the epidemic behaviors on the network. In this paper, we use the number of edges involved in a MPC to measure its size. The relationship between standalone MPCs ($M_1$, $M_2$, …, $M_n$) and overlay MPC ($M_O$) is critical in understanding how overlay diversity affects the functioning of overlay networks.

Many studies have related diversity to the stability of complex systems (Loreau, 2000); hence we also test the relationship between overlay diversity and spreading stability. Here, spreading stability is defined as the capacity of overlay networks to maintain the size of MPC under perturbations, that is, some standalone networks are removed. We evaluate the effects of two perturbation scenarios: random and targeted immunizations. The former scenario randomly chooses a set of standalone networks to remove, and the latter removes standalone networks as the descending sequence of their MPC. We can quantify the relationship between removal ratio, $f$, and overlay MPC, and illustrate how the diversity effect influences the stability of overlay networks.

Moreover, the observed responses of diversity effect can be generated by a combination of two different effects: a 'selection effect' and a 'complementarity effect' (Fig. 2) (Loreau and Hector, 2001; Jiang et al., 2008). The selection effect is based on Price's general theory of selection: selection occurs when changes in $M_O$ are non-randomly related to $M_1, M_2, …, M_n$. Accordingly, selection is measured by a covariance function as in the Price equation of evolutionary genetics. The complementarity effect measures any average change in the $M_O$, whether positive or negative. The sum of these two effects is the net diversity effect; it measures the deviation of $M_O$ from its expected value $M_E$ under the null hypothesis that there is no diversity effect (i.e., diffusion on a randomly choosing standalone network and $M_O$ is the simple average of $M_k$). The additive partition is based on calculating the contribution of each standalone network to form the overlay MPC (Loreau and Hector, 2001). For each $e_{ij} \in$ MPC of the overlay network, we can find a set of standalone networks $\{\Gamma_{k1}, \Gamma_{k2}, …, \Gamma_{km}\}$, which contains all standalone networks with $e_{ij}$. If $M_{kx} > M_{ky}$ ($y=1, 2, …, x-1, x+1, …, m$), then we increase the contribution value of $\Gamma_{kx}$ by 1, that is, $T_{kx} = T_{kx} + 1$, and

$$\sum_{k=1}^{n} T_k = M_O \qquad \text{Eq. 1}$$

In this way, we assigned each edge in the MPC of the overlay network to a standalone network, and $T_k$ indicates the number of contribution edges from standalone network $k$. Hence, the additive partition unifies and relates in a single equation based on the calculation of $T_k$:

$$\begin{aligned}
\Delta M &= M_O - M_E \\
&= \sum_k \frac{T_k}{M_k} M_i - \sum_k \frac{1}{R} M_k \\
&= \sum_k Y_{O,k} M_k - \sum_k Y_{E,k} M_k \quad \text{Eq. 2} \\
&= \sum_k \Delta Y_k M_k \\
&= N \langle \Delta Y \rangle \langle M \rangle + N \operatorname{cov}(\Delta Y, M)
\end{aligned}$$

where $Y_{O,k}$ and $Y_{E,k}$ are the relative contributions of network $k$ in observation and expectation, and $\langle \rangle$ is mean value of data. In this equation, $N\langle \Delta Y \rangle \langle M \rangle$ measures the complementary effect, $N\operatorname{cov}(\Delta Y, M)$ measures the selection effect, and the ratios of selection and complementarity are $N\langle \Delta Y \rangle \langle M \rangle / \Delta M$ and $N\operatorname{cov}(\Delta Y, M)/\Delta M$, separately (Loreau and Hector, 2001). If the diversity effect is mainly attributed to the selection effect, the epidemic behavior is controlled by a few standalone networks with largest $M_k$, and immunization strategies aimed at the networks will be feasible. Contrarily, if the complementarity effect is greater than the selection effect, epidemic behavior is affected by cooperation of all standalone networks, and immunization strategies should consider more about such cooperation behaviors (Fig. 2). Therefore, the additive part of the diversity effect could quantify the mechanism of epidemic behaviors beyond overlay processes.

In sum, the diversity analysis framework not only statistically describes overlay processes of multiple networks, but also quantitatively reveals the mechanism (selection or complementarity) of functioning and stability on multiple overlay networks.

*Heterogeneity of trade networks*

The 83 standalone trade networks connect 202 nodes with 1 to 1893 edges (accounting for 95% of the worldwide vegetables and fruits trade quantity). These standalone trade networks are highly heterogeneous in the connectivity pattern. The probability distribution that a standalone trade network $k$ has $h_k$ edges is exponential and has very large statistical fluctuations (Fig. 3a). Moreover, we also found that the distribution of $M_k$ is also heterogeneous (Fig. 3b). The presence of heterogeneous distributions indicates a possible major impact of a few standalone trade networks with highest connectivity in spreading fatal pathogens. The relationship between heterogeneous distribution and functioning of complex networks has been broadly validated in both theoretical and empirical studies (Clauset et al., 2009). Therefore, with our diversity analysis framework we test the hypothesis that only a few standalone trade networks dominate epidemic behaviors in overlay trade networks.

*Diversity effect*

We applied this framework to 2000 randomly formed overlay trade networks, based on the candidate set of 83 standalone trade networks. The overall diversity effect for overlay trade networks was similar in richness and evenness: a linear increase in average $M_O$ with overlay richness and a linear decrease with overlay evenness (Fig. 4a, b). It means that the number and difference of overlay networks could increase the capacity for spreading pathogens in the overlay trade network. Moreover, the net diversity effect, $\Delta M$, was positive (the grand mean was significantly different from zero) and increased significantly with overlay diversity.

The two components of this net diversity effect, selection and complementarity, had strikingly different performances. The selection effect was minor, and even with a negative value in low overlay richness and evenness (Fig. 4c, d). The negative selection effect operates where the standalone networks with largest $M_k$ do not mainly contribute to the MPC of the overlay network. In contrast, the complementarity effect under any overlay richness and evenness was positive (Fig. 4e, f). Overall, both effects increase with overlay diversity; however, the selection effect slowly increased with overlay richness and evenness (Fig. 4c-f). For example, the average selection effect in $R$=20 only increases ~100 from $R$=2 (i.e., enlarging overlay MPC with about 100 edges). In the same situation, the average complementarity effect increases from ~10 to more than 200. Therefore, the complementarity effect increased more quickly than the selection effect with overlay richness and evenness. Moreover, the complementarity effect is the major contributor to the net diversity effect, and the contribution ratio exceeded 3/4 for all overlay networks; contrarily, the grand mean contribution of selection effect to net diversity effect is no more than 1/4 and the ratio was unaffected by overlay richness and evenness (Fig. 4g, h). The secondary position of selection effect rejected our previous hypothesis based on the heterogeneity of overlay networks. Although there are only a few standalone trade networks with high connectivity, they do not dominate the epidemic behaviors on overlay trade networks. The result is counterintuitive to complex systems with heterogeneous distributions (Albert et al., 2000; Salathe and Jone, 2010). Thus, the high ratio of complementarity suggested that the epidemic behaviors on overlay trade networks were dependent on cooperation among all standalone trade networks, that is, each standalone network, whether with high or low connectivity, will contribute to the formation of MPC in the overlay network (Gu et al., 2011). Such cooperation mechanisms supported the general complex theory that the relationship among components is more important than the components themselves.

Although the contribution ratio of complementarity is relatively unaffected by overlay richness and evenness, the ratio was variable for certain levels of diversity, ranging from ~0 to more than 1 (attributed to a negative selection effect) when overlay richness is small (Fig. 4g). The ratio was also capricious under large overlay diversity, but the value held roughly steady above 1/2. The variable contribution ratio implied the importance of the actual composition, that is, which standalone networks are used to form the overlay network. Thus, we should carefully consider the overlay networks to diffuse real-world pathogens (Table 1). Based on real outbreak data in the United States from 1996 to 2008 from CSPINET (http://www.cspinet.org/), we identified the set of standalone trade networks for 7 major pathogens (including 4 bacteria, 2 viruses, and 1 worm) to represent their potential vectors (belonging to vegetables or fruits). The richness of 7 overlay networks ranges from 5 to 21, and the evenness ranges from 23 to 400. Except for *Clostridium perfringens* and noroviruses, the spread of 5 other pathogens mainly relies on the complementarity effect (Table 1). It supported our conclusion from simulation studies. Selection and complementarity are equally important to the spread of noroviruses (Table 1). Moreover, *Clostridium perfringens* was more dependent on the selection effect. However, we found that the standalone trade networks with large $M_k$ involving *Clostridium perfringens* (e.g., the trade of potato, beans, and tomato) also broadly exist in the overlay networks of other pathogens (Table 1 and *supplementary information*). Thus, the actual compound mode and complex relationship among the standalone networks also play a critical role in controlling epidemic behaviors. It is suggested that the detail of 'cooperation mechanism' among standalone networks dominates the epidemics on overlay trade networks.

*Diversity and stability*

A large number of studies related diversity of complex systems to their stability. To address the stability of the overlay networks in spreading pathogens, we studied changes in the size of the overlay MPC when a fraction of standalone trade networks is removed from the candidate set to form overlay networks. In the random immunization scenarios, we randomly chose standalone networks to remove, and found that average $M_O$ remains unchanged under increasing fractions of removal when richness is constant (Fig. 5a). Even when as many as 10 standalone networks fail, the epidemic behavior in the overlay network is unaffected. This stability of overlay trade networks is rooted in the dominance of complementarity, because the complementarity implied that cooperation rather than special standalone networks plays a critical role in controlling epidemics. Random removal does not alter cooperation among standalone networks statistically, and hence has no impact on the overlay networks (Albert et al., 2000).

In the second scenario, to simulate a targeted immunization, we first removed the standalone network with the largest MPC, and continued selecting and removing networks in descending order of their MPC. When the most connected standalone networks are deleted, the average overlay MPC with the same richness decreases rapidly (Fig. 5a). The phenomenon is not different with complex systems with heterogeneous distribution. These systems are always unstable to targeted immunization (Albert et al., 2000; Salathe and Jone, 2010). However, we found that the complementarity effect could largely increase the stability of overlay trade networks. When faced with targeted immunization, for overlay networks with low $R$, a relatively small increase in overlay richness could compensate for the failure of high-fraction standalone networks with a large MPC (Fig. 5b). For example, when the top 10 standalone networks are removed, the average $M_O$ decreases to ~1/2 of unperturbed systems when $R$=5. However, if we raise the average richness by 4.7 for perturbed systems, the average MPC could equal that of unperturbed systems. For overlay networks with high $R(>10)$, complementarity could also compensate for the loss of standalone networks with the largest MPC, but the increase in overlay richness will exceed the number of standalone networks removed. Thus, the complementarity effect overcomes targeted immunization strategies and enhances the stability of epidemics on overlay trade networks. In sum, the targeted immunization will not be as effective in overlaying trade networks as its performance in other complexes with heterogeneous distribution. In order to control the epidemic on overlay trade networks, we should pay more attention to the cooperation mechanism rather than a few standalone trade networks with the largest MPC.

**Conclusion**

In this paper, we extended diversity theory from biology research to study the overlay of multiple heterogeneous networks. The diversity analysis could describe not only the relationship between the feature of standalone networks and the resulting overlay networks, but also revealed the mechanism beyond overlay processes. By applying diversity analysis to global trade networks, we proved the practical and theoretical value of our framework. Our work is an initial step towards finding general tools to analyze overlay networks. In the future, by adjust the parameters in current framework (e.g. defining other functions rather than percolation cluster, changing metrics of richness or evenness, refining the partition of net diversity effect), the framework could extend to other complex systems consisted of multiple networks, including multiplex networks and interdependent networks (Buldyrev et al., 2010; Parshani et al., 2010a; 2010b; Gao et al., 2012; Paczuski et al., 2012).

1 **Figures and tables**

2 **Figure 1.**

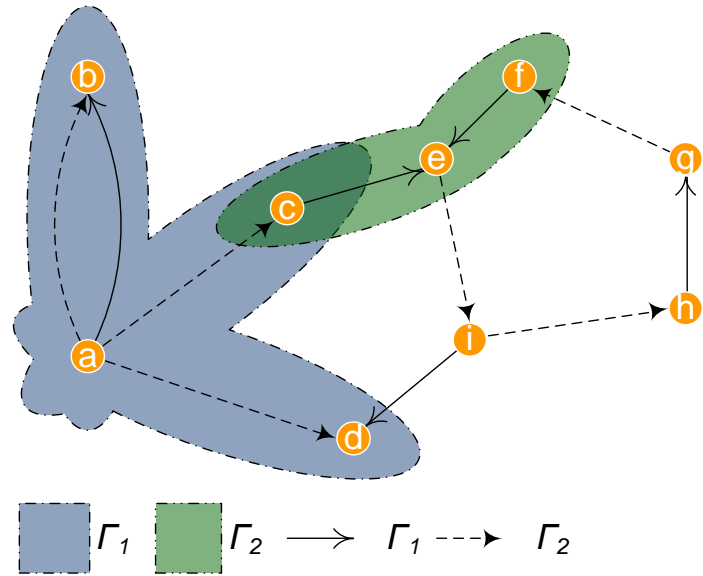

4 **Figure 1. Illustration of overlay processes.** The figure demonstrates the overlay of

5 two standalone networks $\Gamma_1$ and $\Gamma_2$. Both networks cover all 9 nodes, the dash lines

6 belong to $\Gamma_1$, and the solid lines belong to $\Gamma_2$. The maximal percolating clusters of $\Gamma_1$

7 and $\Gamma_2$ is indicated by the blue and green background respectively. The maximal

8 percolating cluster of the overlay network covers all edges.

**Figure 2.**

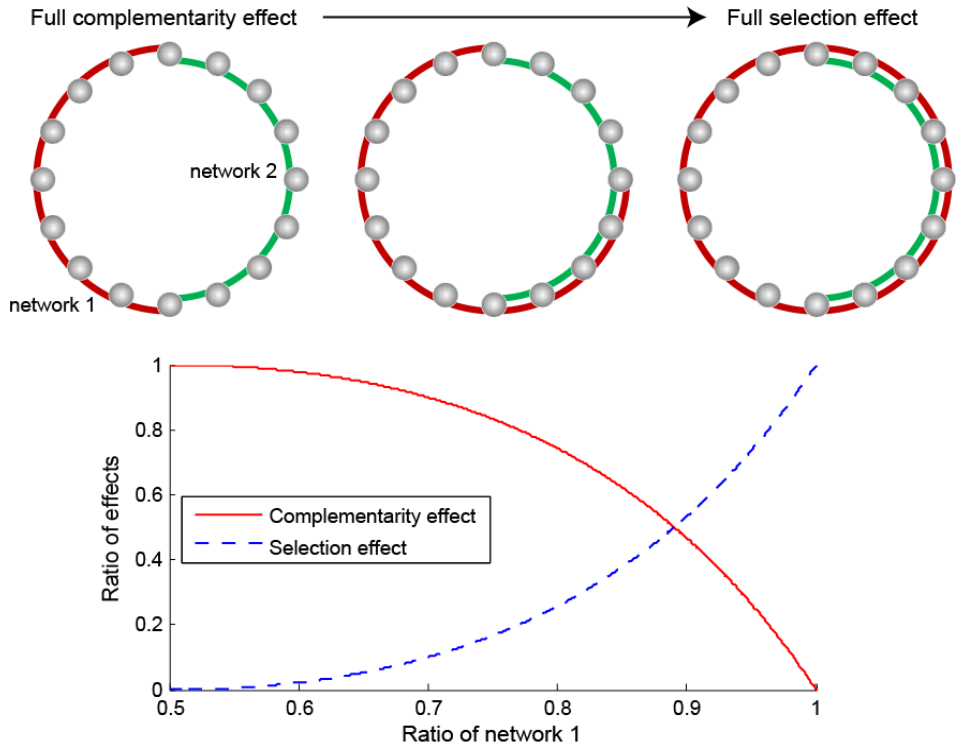

**Figure 2. Illustration of complementarity and selection effect.** The above part of figure uses two cycle networks to demonstrate the complementarity and selection effect of overlay networks. The network 2 is fixed to the half part of the cycle. If two networks do not have common edges (the network 1 covers the other half of the cycle), the net overlay effect is completely attributed to complementarity effect. If the network 1 covers the full cycle, the net overlay effect is completely attributed to selection effect. The bottom part of figure shows the change of two effects along with the increase of network 1. The simulating cycle network consists of 1000 nodes.

**Figure 3.**

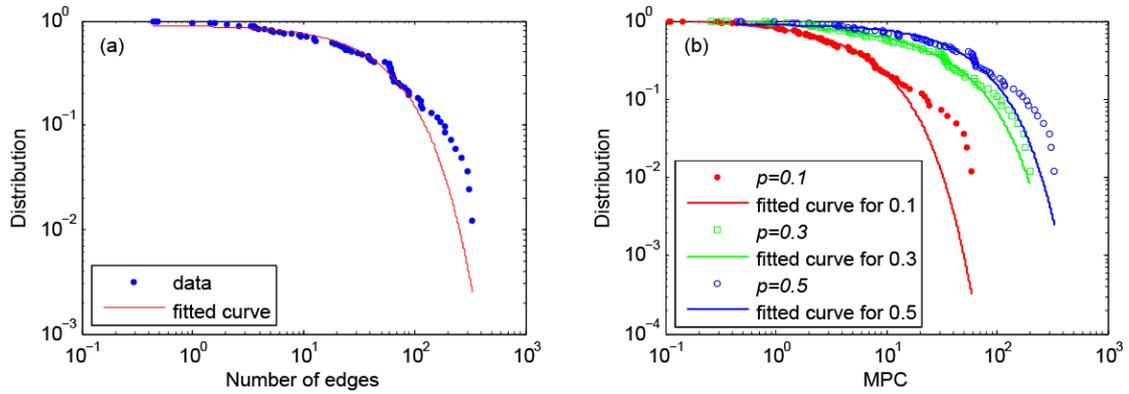

**Figure 3. Heterogeneous distribution of 83 standalone trade networks.** (a) The log-log plot of the distribution of edge number. (b) The log-log plot of the distribution of maximal percolating cluster under different percolation probability $p$. All four data are fitted by exponential curve, and coefficient of determination $R^2>0.8$.

**Figure 4.**

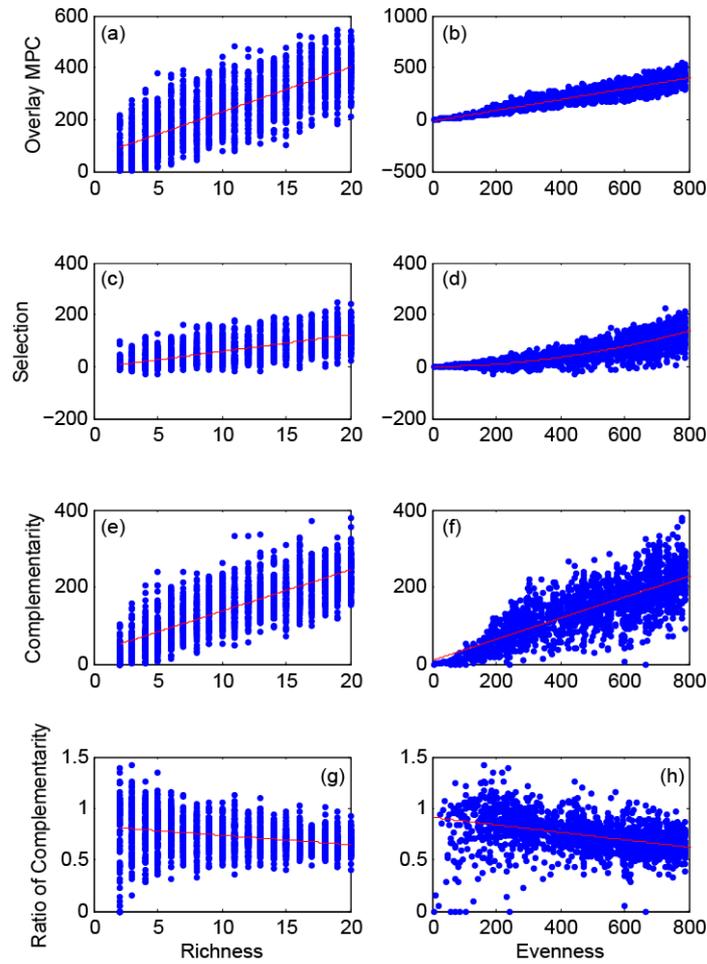

**Figure 4. Diversity effect of overlay trade networks.** (a)-(g) the relationship between overlay richness or evenness and overlay MPC, selection effect, complementarity effect or ratio of complementarity effect. Except for (d), other 7 data pairs are fitted by linear curves, and (d) is fitted by power curve. The coefficient of determination of (a)-(f), $R^2>0.6$, but $R^2$ of (g) and (h) is less than 0.2. In the figure, we only show the ratio of complementarity effect because the sum of ratio of selection and complementarity is equal to 1. The calculation of MPC in the figure is based on percolating probability $p=0.3$.

**Figure 5.**

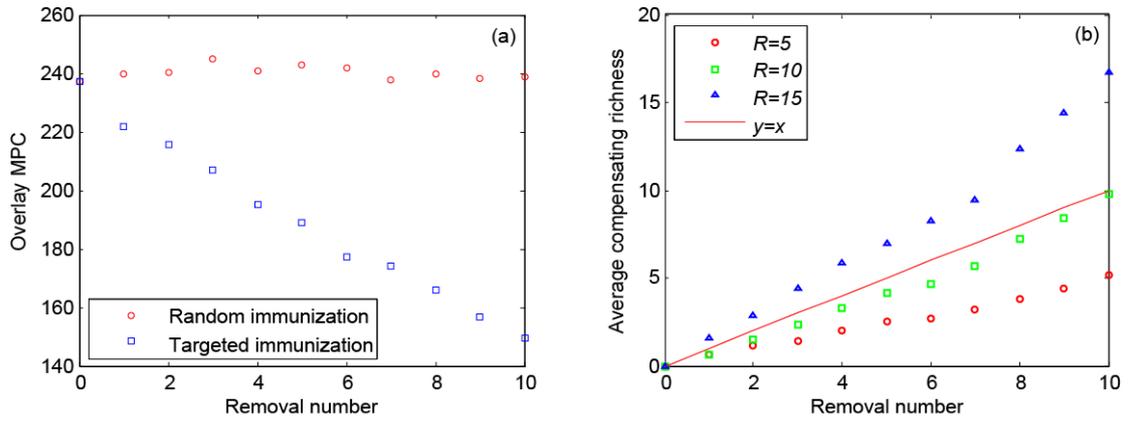

**Figure 5. Stability of overlay trade networks.** (a) The average overlay MPC changes with random and targeted immunization strategies. The overlay richness is 10, and the simulation generates 2000 different overlay trade networks for each removal number. (b) The compensating richness for offsetting targeted immunization. The simulation performs on different original overlay richness ($R$=5, 10, 15), and generates 2000 different overlay trade networks for each removal number.

Table 1.

| Type | Pathogen | $R$ | $S$ | $RS$ | $C$ | $RC$ |
|---|---|---|---|---|---|---|
| Bacteria | Salmonella | 21 | 205.10 | 0.44 | 266.18 | 0.56 |
| Bacteria | Campylobacter jejuni | 8 | 110.77 | 0.37 | 189.60 | 0.63 |
| Bacteria | Escherichia coli | 14 | 168.84 | 0.49 | 175.37 | 0.51 |
| Bacteria | Clostridium perfringens | 12 | 194.45 | 0.59 | 137.13 | 0.41 |
| Worm | Cyclospora cayetanensis | 5 | 30.05 | 0.39 | 46.15 | 0.61 |
| Virus | Noroviruses | 21 | 262.13 | 0.50 | 264.77 | 0.50 |
| Virus | Hepatitis A | 6 | 120.58 | 0.47 | 136.26 | 0.53 |

$^*$ $R$ is overlay richness, $S$ is selection effect, $RS$ is the ratio of selection effect, $C$ is complementarity effect, $RC$ is the ratio of complementarity effect.

$^{**}$ The calculation of MPC is based on percolating probability $p=0.3$.